\documentclass[11pt]{article}
\usepackage[paperheight=12in,paperwidth=8.75in]{geometry}
\usepackage{setspace}
\usepackage{setspace}
\usepackage[T1]{fontenc}
\usepackage{times}
\usepackage{palatino}
\usepackage{lmodern}
\usepackage[latin1]{inputenc}
\usepackage{epsfig}
\usepackage[english]{babel}
\usepackage{color}
\usepackage{graphicx}
\usepackage{dcolumn}
\usepackage{amsmath,amssymb,amsfonts}
\usepackage[all]{xy}
\usepackage{cite}
\usepackage[dvipsnames]{xcolor}
\usepackage{physics}
\usepackage{mathrsfs}
\usepackage{authblk}
\usepackage{geometry}
\usepackage{abstract}
\usepackage[english]{babel}
\usepackage[T1]{fontenc}
\usepackage[colorlinks,urlcolor=blue,citecolor=blue]{hyperref}
\usepackage{amssymb,amsmath}
\usepackage{bm}%
\usepackage{stmaryrd}
\usepackage{comment}
\usepackage{datetime}

\def\nn{\nonumber}

\def\hf{\frac{1}{2}}

\def\Zgr{$ {\mathbb Z}_2$}
\def\Z2{${\mathbb Z}_2 \times {\mathbb Z}_2$}
\def\g{\mathfrak{g}}

\begin{document}
\numberwithin{equation}{section}
\newcommand{\boxedeqn}[1]{%
  \[\fbox{%
      \addtolength{\linewidth}{-2\fboxsep}%
      \addtolength{\linewidth}{-2\fboxrule}%
      \begin{minipage}{\linewidth}%
      \begin{equation}#1\end{equation}%
      \end{minipage}%
    }\]%
}


\newsavebox{\fmbox}
\newenvironment{fmpage}[1]
     {\begin{lrbox}{\fmbox}\begin{minipage}{#1}}
     {\end{minipage}\end{lrbox}\fbox{\usebox{\fmbox}}}

\raggedbottom
\onecolumn

\begin{center}
{\Large \bf
The $osp(1|2)$ \Z2 graded algebra and its irreducible representations \\
}
\vspace{6mm}
{Fahad Sameer Alshammari${}^1$,  Md Fazlul Hoque${}^{2,4}$ and Jambulingam Segar${}^3$
}
\\[6mm]
\noindent ${}^{1}${\em 
Department of Mathematics, College of Science at Alkharj, Prince Sattam bin Abdulaziz University, Alkharj, 11942, Saudi Arabia }
\\[3mm]
\noindent ${}^{2}${\em
Department of Physics, Czech Technical University in Prague, Faculty of Nuclear Sciences and Physical Engineering, B\v{r}ehov\'a 7, 115 19 Prague 1, Czech Republic} 
\\[3mm]
\noindent ${}^{3}${\em
Department of Physics, Ramakrishna Mission Vivekananda College, Mylapore, Chennai 600 004, India} 
\\[3mm]
\noindent ${}^{4}${\em
Pabna University of Science and Technology, Faculty of Science, Department of Mathematics, Pabna 6600, Bangladesh }\\ 
[1mm]
\vspace{4mm}
{\footnotesize Email: fazlulmath@pust.ac.bd; segar@rkmvc.ac.in }
\end{center}

\vskip 1cm
\begin{abstract}
\noindent In this paper, we construct finite dimensional irreducible representations for two different versions of \Z2 graded $osp(1|2)$ algebra based on eight- and ten-generators. We find that there are two different second-order Casimir operators for the ten generators version of the algebra  one corresponding to the $(0,0)$ sector and another in the $(1,1)$ sector. Consequently, it is shown that the eight-generator version of the algebra has only one Casimir invariant in the $(0,0)$ sector. We present the differential operator realizations of these algebras. Starting with the highest weight, we construct the states of the irreducible finite-dimensional representations for both versions of the algebras. The matrix realizations of the generators on the representation space corresponding to these states are written down explicitly.
\end{abstract}

\section{Introduction}
This study aims to construct finite dimensional irreducible representations of the \Z2-graded extension of $osp(1|2)$ superalgebra based on eight- and ten-generators, find their corresponding quadratic Casimir invariants in various sectors and various finite dimensional irreducible representations. 

It is well known that Lie algebras and Lie superalgebras play a crucial role in the theoretical understanding of physical systems. In fact, these algebraic structures help us to characterize the systems completely in many cases. The Lie superalgebra can be viewed as an $Z_2$ extension of the Lie algebras. But the Lie algebraic structures can be extended over an abelian group. The Lie algebra inherits an extended graded structure according to the Abelian group over which it is extended. 

Let us recall that color superalgebras were first introduced by Rittenberg and Wyler \cite{rw1, rw2} in the 1970s,  though the related structures were studied in earlier work of Ree \cite{Ree1960,Sch1979}. These are a generalization of Lie superalgebras.  The color superalgebra of \Z2-graded has been investigated from both physical and mathematical perspectives; for example, the articles\cite{Zha2003, aktt1, NaPsiJs, NA2 } and references therein. Models for superconformal mechanics are also studied to generalize the characterizations of the \Z2-graded superconformal algebras with their extensions. \cite{AzKaDo}. 

It can be recalled that, $osp(1|2)$ plays the same role in the study of superalgebras as that of the role played by $sl_2$ in ordinary Lie algebras. In order to understand role of symmetry algebra better and its applications, a good knowledge of the representation theory
of symmetry algebra is essential. Very few works are available on the representation theory of \Z2 graded algebras. This provides the motivation for this work. 

A key role is played by the Casimir operator for $osp(1|2)$ in representation theory, physical applications, the exact analysis of integrable quantum Hamiltonian systems, and the interplay between the two. For instance, they are of particular importance in the dynamical symmetry algebra approach whereby an algebraic quantum Hamiltonian can be written in terms of Casimir operators corresponding to a chain of Lie subalgebras. Among the best-known examples are the interacting boson model in nuclear physics and the vibron model in molecular physics \cite{Dee85, Fil81, Iac80}. 

In this work, section 2 recalls the basic definitions of \Z2 graded algebras, section 3 derives the quadratic Casimir and the differential operator realizations of the algebra,  section 4 discusses the characterization of finite dimensional irreducible representations
and an example case if worked out in detail, in section 5 we discuss all the above results for another version of \Z2 graded $osp(1|2)$ algebra with eight generators and we end the paper with a summary and conclusions.

\section{Definitions and relevant structures}

\subsection{Color superalgebra of \Z2 grading}
At the beginning we briefly recall the definition of \Z2-graded color superalgebra. Let us consider a vector space $\g$ and an element $ \bm{a} = (a_1, a_2)$ of \Z2. The direct sum of graded components of $\g$ is defined by the decompositions 
\begin{equation}
   \g = \bigoplus_{\bm{a}} \g_{\bm{a}} = \g_{(0,0)} \oplus \g_{(0,1)} \oplus \g_{(1,0)} \oplus \g_{(1,1)}.
\end{equation}
Then $\g$ is a \Z2-graded color superalgebra together with a bilinear operation, denoted by 
$ \llbracket \cdot, \cdot \rrbracket, $ the general Lie bracket, satisfying the identities \cite{rw1,rw2}:
\begin{align}
  & \llbracket X_{\bm{a}}, Y_{\bm{b}} \rrbracket \in \g_{\bm{a}+\bm{b}}  \text{(grading)},
  \\[3pt]
  & \llbracket X_{\bm{a}}, Y_{\bm{b}} \rrbracket = -(-1)^{\bm{a}\cdot \bm{b}} \llbracket Y_{\bm{b}}, X_{\bm{a}} \rrbracket \text{(graded antisymmetry)},
    \\[3pt]
  & \llbracket X_{\bm{a}}, \llbracket Y_{\bm{b}}, Z_{\bm{c}} \rrbracket \rrbracket
    = \llbracket \llbracket X_{\bm{a}},  Y_{\bm{b}}\rrbracket, Z_{\bm{c}} \rrbracket 
    + (-1)^{\bm{a}\cdot\bm{b}} \llbracket Y_{\bm{b}}, \llbracket X_{\bm{a}}, Z_{\bm{c}} \rrbracket \rrbracket  \text{(graded Jacobi identity)},
    \label{gradedJ}
\end{align}
where the vectors are defined by
\begin{equation}
  \bm{a} + \bm{b} = (a_1+b_1, a_2+b_2) \in {\mathbb Z}_2 \times {\mathbb Z}_2, \qquad \bm{a}\cdot \bm{b} = a_1 b_1 + a_2 b_2,
\end{equation}
and $ X_{\bm{a}}, Y_{\bm{b}}, Z_{\bm{c}} $  are homogeneous elements of $ \g_{\bm{a}}$. The \Z2-graded unital associative algebra performed as the enveloping algebra of $\g$ with the relations
\begin{equation}
 \llbracket X_{\bm{a}}, Y_{\bm{b}} \rrbracket =  X_{\bm{a}} Y_{\bm{b}} - (-1)^{\bm{a}\cdot \bm{b}}
Y_{\bm{b}} X_{\bm{a}}. \label{gradedcom}
\end{equation}
The expression such as $ X_{\bm{a}} Y_{\bm{b}}$ is understood to denote the associative product
on the enveloping algebra. 
In other words, by definition, in the enveloping algebra the general Lie bracket $ \llbracket \cdot, \cdot
\rrbracket $ for homogeneous elements coincides with either a commutator or anticommutator. 

 This is a natural generalization of Lie superalgebra which is defined on \Zgr-grading structure:
\begin{equation}
  \g = \g_{(0)} \oplus \g_{(1)}
\end{equation}
with 
\begin{equation}
  \bm{a} + \bm{b} = (a+b), \qquad \bm{a} \cdot \bm{b} = ab.
\end{equation}
It is remarked that $ \g_{(0,0)} \oplus \g_{(0,1)} $ and $ \g_{(0,0)} \oplus \g_{(1,0)} $ are sub-superalgebras of the \Z2-graded superalgebra $\g$.

\subsection{The general linear superalgebra}
We consider  general linear superalgebra $gl(m_1,m_2|n_1, n_2)$ \cite{PhJo}, where the labels $m_1$, $m_2$,
$n_1$ and $n_2$ are non-negative integers. As in the $\mathbb{Z}_2$-graded case, a convenient basis is the standard one comprising (homogeneous) elementary
matrices $E_{i\,j}$ with 1 in the entry of row $i$, column $j$, and 0 elsewhere. To describe the grading of these basis elements, we use the graded index
\begin{equation}
\bar {d}_i = \left\{
\begin{array}{ll}
(0, 0); & i = 1, . . . ,m_1\\
(1, 1);  & i = m_1 + 1, . . . ,m_1 + m_2 \\
(1, 0), & i = m_1 + m_2 + 1, . . . ,m_1 + m_2 + n_1 \\
(0, 1); & i = m_1 + m_2 + n_1 + 1, . . . ,m_1 + m_2 + n_1 + n_2,
\end{array}
\right.
\end{equation}
where we adopt the convention that in a situation where one of the labels $m_1$, $m_2$, $n_1$ or $n_2$
 is zero, the case $i = j, . . . , j-1$ does not exist.
We then define $\bar {d}_{i\,j} \equiv \bar {d}(E_{i\,j}) =\bar{d}_i +\bar{d}_j$, where the sum is taken component-wise, modulo 2, as described earlier. The \Z2 -graded Lie product
then takes on the form
\begin{equation}
\llbracket  E_{ij} ,E_{kl} \rrbracket = \delta_{jk}E_{il} - (-1)^{\bar{d}_{ij}\cdot\bar{d}_{kl}}\delta_{il} E_{kj}.
\end{equation}

\section{The Casimir invariants and differential realisation}
The \Z2 graded algebra $osp(1|2)$ was introduced in \cite{rw2}, we denote this algebra by $\g$. The Verma modules of this algebra are studied thoroughly in \cite{AzKa}. We follow the notations in \cite{AzKa}.
As a vector space, it has the  basis elements 
$\{ R, L_{\pm},a_{\pm}, \tilde{R}, {\tilde{L}}_{\pm}, {\tilde{a}}_{\pm} \}$ with four subspaces  labelled by the elements of \Z2:
\begin{equation}
R, L_{\pm} \in \g_{(0,0)},\quad a_{\pm} \in \g_{(0,1)}, \quad {\tilde{a}}_{\pm} \in \g_{(1,0)}, 
\quad \tilde{R}, {\tilde{L}}_{\pm} \in \g_{(1,1)}.
\end{equation}
The elements of the algebra satisfy the following non-vanishing identities:
\begin{equation}
\begin{array}{llll}
 [R,L_{\pm}] = \pm 2 L_{\pm}, & [R, \tilde{L}_{\pm}] = \pm 2\tilde{L}_{\pm},& [R, a_{\pm}] = \pm a_{\pm}, & 
 [ R, {\tilde{a}}_{\pm}] = \pm {\tilde{a}}_{\pm}, \\[3pt]
[\tilde{R},L_{\pm}] = \pm 2\tilde{L}_{\pm}, & [\tilde{R}, \tilde{L}_{\pm}] = \pm 2L_{\pm},& 
\{\tilde{R}, a_{\pm}\} = \tilde{a}_{\pm}, & \{\tilde{R}, \tilde{a}_{\pm}\} = a_{\pm}, \\[3pt]
[L_{+},L_{-}] = -R, & [L_{\pm}, \tilde{L}_{\mp}] = \mp \tilde{R}, & [\tilde{L}_{+}, \tilde{L}_{-}] = -R, & \\[3pt]
[L_{\pm}, \tilde{a}_{\mp}] = \pm \tilde{a}_{\pm},& [L_{\pm}, a_{\mp}] = \mp a_{\pm}, &
 \{ \tilde{L}_{\pm}, a_{\mp}\} = -\tilde{a}_{\pm}, & \{\tilde{L}_{\pm}, \tilde{a}_{\mp}\} = a_{\pm},\\[3pt]
\{a_{+}, a_{-}\} = 2R, & [a_{\pm}, \tilde{a}_{\mp}] = \pm 2 \tilde{R},& \{ \tilde{a}_{-}, \tilde{a}_{+} \} = 2R, & \\[3pt]
[a_{\pm}, \tilde{a}_{\pm}] = \mp 4 \tilde{L}_{\pm}, & \{a_{\pm}, a_{\pm}\} = 4L_{\pm},& 
\{\tilde{a}_{\pm}, \tilde{a}_{\pm}\} = -4L_{\pm}.&
\end{array}
\label{Z21}
\end{equation}
A Cartan subalgebra of $\g$ can be constructed  in terms of maximally nilpotent subalgebra of $\g$, it is, $\mathfrak{h} := \text{lin.span}\{R,\tilde{R}\}$. The Cartan subalgebra is a subspace of $\g_{(0,0)}\oplus \g_{(1,1)}$. It can be seen that there is a problem in defining the triangular decomposition with respect to $\mathfrak{h}$. This problem is overcome, see \cite{AzKa} and the triangular decomposition is introduced only with respect to degree $(0,0)$ element. The elements of $\g$ and their eigenvalue for the $adR$ are given below:
\begin{equation}
\begin{array}{l|llll}
 &(0,0)&(0,1)&(1,0)&(1,1) \\[5pt] 
 \hline
 +2&L_{+}& &  & \tilde{L}_{+} \\[5pt]
 +1& & a_{+}&\tilde{a}_{+}& \\[5pt]
 0&R & &     & \tilde{R}  \\[5pt]
 -1&&a_{-}&\tilde{a}_{-} & \\[5pt]
 -2&L_{-}&& & \tilde{L}_{-}
 \end{array}
 \label{Triandec}
 \end{equation}
A Casimir operator is defined as an element of the centre of the universal enveloping algebra of a Lie algebra. It is typically a polynomial expression in the basis element, whose bracket
with all the generators in the basis of the algebra is zero. It plays a very important role in the representation theory of Lie algebras and in the analysis of integrable quantum Hamiltonian systems. 
The notion of Casimir element finds suitable generalisation according to the structural extensions of the Lie algebras\cite{Fa}. The role played by the Casimir operator in the representation theory
of supersymmetric $osp(1|2)$ is  well known \cite{Caosp12}.

We investigate quadratic Casimir invariant of the above algebra $\g$.  We start with the quadratic case considering a general expression of the form 
\begin{equation}
C_2=\alpha R + \beta R^2 + \gamma \tilde{R}^2 +\delta L_{+}L_{-}+ \lambda \tilde{L}_{+}\tilde{L}_{-} 
+\rho a_{+}a_{-} + \sigma \tilde{a}_{+}\tilde{a}_{-}, \\
\end{equation}
in the $(0,0)$ sector with arbitrary constants $\alpha, \beta, \gamma, \delta, \lambda, \rho, \sigma$  to be determined. Demanding that $C_2$ commutes with the elements of $\g$ leads to a set of
algebraic relations among $\alpha, \beta, \gamma, \delta, \lambda, \rho, \sigma$, which is solved to get the relations
\begin{equation}
\beta=-\hf \alpha =\gamma, \,\,\,\,\, \delta=2\alpha=\lambda,\,\, \,\,\, \rho=\hf \alpha = \sigma,
\end{equation}
which leads to the explicit form of the quadratic Casimir invariant in the sector $(0,0)$,
\begin{eqnarray}
{C_2}^{(0,0)}&=& R-\hf R^2 -\hf \tilde{R}^2 + 2L_{+}L_{-} + 2\tilde{L}_{+}\tilde{L}_{-}+\hf a_{+} a_{-}
+\hf \tilde{a}_{+}\tilde{a}_{-},\, \\[3pt]
&=&-R-\hf R^2 -\hf \tilde{R}^2 +2 L_{-}L_{+} +2 \tilde{L}_{-}\tilde{L}_{+}-\hf a_{-}a_{+}-\hf \tilde{a}_{-}\tilde{a}_{+}.
\end{eqnarray}
Using the same techniques, one can find another Casimir in the $(1,1)$ sector as
\begin{equation}
{C_2}^{(1,1)}= \mu(L_{+}\tilde{L}_{-} + \tilde{L}_{+}{L}_{-}+\frac{1}{4}a_{+} \tilde{a}_{-}
+\frac{1}{4} \tilde{a}_{+}{a}_{-}+\hf \tilde{R}-\hf R \tilde{R}) + \nu,
\end{equation}
where $\mu$ and $\nu$ are arbitrary constants. We will see how the eigenvalues of these Casimir operators label the states in the representation space.
\noindent
We now construct differential realizations. The representation space of $\g$ is a \Z2 graded vector space $V$ with the actions of the elements in $EndV$. Starting from a
particular vector  ${| \,\,\, \rangle} \, \in V$, the other states in the vector space is obtained by the suitable action of:
\begin{equation}
e^{xL_{-}}e^{\theta a_{-}} e^{\psi {\tilde{a}}_{-}}e^{z \tilde{L}_{-}}.
\end{equation}
On ${|\,\,\,\,\rangle} $, with:
\begin{equation}
x \in (0,0), \,\,\, \theta \in (0,1), \,\,\, \psi \in (1,0), \,\,\, z \in (1,1)
\end{equation}
and parameters $r \in (0,0)$, $\tilde{r}\in(1,1)$ and using the defining relations of the algebra, this produces the following linear differential realization of the generators of the algebra (\ref{Z21})
\begin{eqnarray}
L_{-}&=&\partial_x, \\
\tilde{L}_{-}&=& \partial_z, \\
a_{-}&=&\partial_{\theta}+2\theta \partial_x, \\
\tilde{a}_{-}&=&\partial_{\psi}-2\psi \partial_x +4\theta \partial_z, \\
R&=&r+2x \partial_x + 2 z \partial_z + \theta \partial_{\theta} + \psi \partial_{\psi}, \\
\tilde{R}&=& \tilde{r}+2z\partial_x + 2x\partial_z +\theta\partial_{\psi}+\psi \partial_{\theta}-
4\psi \theta \partial_x, \\
a_{+}&=&2\theta r - 2 \psi \tilde{r}+2\theta x \partial_x + 2z\psi \partial_x + 2\theta \psi \partial_{\psi}-z\partial_{\psi} +x\partial_{\theta}-4z\theta\partial_z,\\
\tilde{a}_{+}&=&2\psi r -2\theta \tilde{r}+2z\theta\partial_x+2\psi x \partial_x +z\partial_{\theta}-2\theta\psi \partial_{\theta}-x \partial_{\psi}-4\theta x \partial_z, \\
L_{+}&=&x r +z\tilde{r}-2\theta \psi \tilde{r}+x^2 \partial_x +z^2 \partial_x -4z \theta \psi \partial_x +x \psi \partial_{\psi}+x\theta \partial_{\theta}\nn \\&& +2zx\partial_z + z\theta \partial_{\psi}+z\psi \partial_{\theta}, \\
\tilde{L}_{+}&=&zr+x\tilde{r}-2\theta \psi r -4\theta \psi x \partial_x=2zx\partial_x +z\theta\partial_{\theta} +\psi x \partial_{\theta}+z\psi \partial_{\psi}\nn \\&& +\theta x \partial_{\psi}+z^2\partial_z +x^2\partial_z.
\end{eqnarray}
These differential realizations are expected to play a definitive role in studying the theoretical and mathematical aspects of systems which will be characterised by $\g$. Moreover, it will help us in extending
the algorithms developed in \cite{Fa} to the generalised algebraic structures studied here. One can establish that this realization is irreducible by substituting the above for the generators in the Casimir and showing that it is a multiple of identity operator.

\section{Finite dimensional irreducible representation} 
The triangular decomposition of the algebra $\g$ with respect to the operator $R$ can be written as:
\begin{equation}
\g=\g^{+}\, \oplus \g^{0}\, \oplus \g^{-},
\end{equation} 
with
\begin{equation}
\begin{array}{ll}
\g^{+}&=\{L_{+},a_{+},\tilde{a}_{+},\tilde{L}_{-} \}, \\[3pt]
\g^{0}&=\{ R, \tilde{R} \}, \\[3pt]
\g^{-}&=\{L_{-},a_{-},\tilde{a}_{-}, \tilde{L}_{-} \}.
\end{array}
\end{equation}
We define the highest weight state to be a state in the representation space, to be a state annihilated by $\g^{+}$, that is,
\begin{equation}
X | \,\,\,\rangle = 0 \,\,   \forall X  \in \g^{+}.
\end{equation}

We realize that the representation space of the algebra can be characterized by the eigenvalues of Casimir and $R$. The representation space is a \Z2 graded vector space.
Writing the states with their degree as ${| \,\,\, \rangle}_{a,b} $ and denoting the generators generically by $X_{a,b}$, then the action on the states is defined by
\begin{equation}
X_{a,b} | \,\, {\rangle}_{c,d}=|\,\,\,{\rangle}_{a+c.b+d}.
\end{equation}
Here the addition ${a+c,b+d}$ is subject to the restriction of $mod \,\,2$.

\section{The action of the generators}
We define the highest state vector as follows
\begin{equation}
L_{+} |  \,\,\, \rangle =0, \,\,\,  a_{+} | \,\,\, \rangle =0,\,\,\, \tilde{a}_{+} | \,\,\, \rangle =0.
\end{equation}
The other states in the representation space will be obtained by the action of $L_{-}$, $a_{-}$
and $\tilde{a}_{-}$ on the highest weight state. If we demand that the representation space is finite-dimensional, then the action of generators in $\g^{-}$ will terminate at some finite steps. 
We know that the states in finite-dimensional representation can be completely described by the eigen generators which can be diagonalized simultaneously. The quadratic Casimir and the 
$R$ are such a set of completely diagonalizable generators, denoting their eigenvalues by $p$ and $q$, i.e :
\begin{eqnarray}
C_{2}^{0,0} |p,q \rangle &=& p |p,q\rangle, \\[3pt]
R |p,q\rangle &=& q |p,q\rangle.
\end{eqnarray}
From the algebra $\g$, we may observe the following:
\begin{itemize}
\item The action of $L_{\pm},$ increases (decreases) the eigenvalue $q$ by two, without changing the degree of the state vector. 
\item The action of $a_{\pm}$ increases (decreases) the eigenvalue $q$ by 1, with changing the degree of the state vector. 
\item In a fixed representation corresponding to the highest eigenvalue, say $q_{max}$, the eigenvalues of $R$ range from $q_{max} $ to $q_{min}$ with $q_{max}=-q_{min}$.
\end{itemize}

It can be checked that the eigenvalue of $R^2$ and $\tilde{R}^2$ are the same on the highest weight
vector. Taking the highest weight state to be characterised by the largest eigenvalue of $R$, i.e. $q_{max}$, from now on we denote it by $\ell$. The eigenvalue of $C_2^{0,0}$ on the highest weight
vector is $\ell(\ell+1)$ We write this state as:
\begin{equation}
|\ell; \ell\rangle.
\end{equation}
Here we take the highest weight state as normalized, that is,
\begin{equation}
<\ell;\ell|\ell;\ell>=1.
\end{equation}
From the defining relations of the algebra, one can establish
\begin{eqnarray}
L_{+}L_{-}^n&=&L_{-}^nL_{+}-n(R+n-1)L_{-}^{n-1}\nn \\
&=&L_{-}^nL_{+}+L_{-}^{n-1}[-n(R-n+1)]
\end{eqnarray}
and the following:
\begin{eqnarray}
\langle \,\,\, {L_{+}}^n {L_{-}}^n \,\,\, \rangle &=& (-)^n n!\langle \,\,\, (R-n+1)(R-n+2)(R-n+3)....R \,\,\,\rangle, \\
\langle \,\,\,  {L_{+}}^na_{+} {L_{-}}^n a_{-} \,\,\, \rangle &=& (-)^{n+1}n! \langle \,\,\, (R-n+1)(R-n+2)....R 
a_{-}a_{+} \,\,\,\rangle \\ &&+2(-)^n n!\langle \,\,\, (R-n)(R-n+1)(R-n+2)....R \,\,\,\rangle, 
\\ \langle \,\,  {L^n_{+}}\tilde{a}_{+} {L^n_{-}} \tilde{a}_{-} \,\,\, \rangle &=& (-)^{n+1}n! \langle \,\,\, (R-n+1)(R-n+2)....R \tilde{a}_{-}\tilde{a}_{+} \,\,\,\rangle
\\ && +2(-)^n n!\langle \,\,\, (R-n)(R-n+1)(R-n+2)....R \,\,\,\rangle,
  \\
 \langle \,\,  {L^n_{+}}a_{+}\tilde{a}_{+}{L^n_{-}} \tilde{a}_{-}a_{-} \,\,\, \rangle &=& 
\langle L_{+}^nL_{-}^na_{-}\tilde{a}_{-}a_{+}\tilde{a}_{+}\rangle -2\langle L_{+}^nL_{-}^na_{-}a_{+}R\rangle-2\langle L_{+}^nL_{-}^n\tilde{a}_{-}\tilde{a}_{+}R\rangle \nn \\
&&-2\langle L_{+}^nL_{-}^n \tilde{a}_{-}a_{+}\tilde{R}\rangle-2\langle L_{+}^nL_{-}^n{a}_{-}\tilde{a}_{+}\tilde{R}\rangle - 4\langle L_{+}^nL_{-}^na_{-}a_{+}\rangle \nn \\
&& + 4\langle L_{+}^nL_{-}^n\tilde{a}_{-}\tilde{a}_{+}\rangle +4\langle L_{+}^nL_{-}^nR^2\rangle- 4\langle L_{+}^nL_{-}^nR\rangle- 4\langle L_{+}^nL_{-}^n\tilde{R}^2\rangle.\nn \\
&&
\end{eqnarray}
Making use of the fact that the action of $L_{+}$ on the highest weight is zero, and the action of $R$ gives $\ell$, we get the following result, when we take
the expectation value with respect to the highest weight state.
We find
\begin{equation}
<\ell;\ell|L_{+}^nL_{-}^n|\ell; \ell>=(-1)^n\frac{n! {\ell}!}{(\ell-n)!}.
\end{equation}
In order to make the representations real, we take the adjoint of $L_{-}$ to be
$-L_{+}$, and define the the normalized vector $|\ell; \ell-2n>$ as
\begin{equation}
|\ell; \,\,\ell-2n>=\sqrt{\frac{(\ell-n)!}{n!\,\,\ell!}} L_{-}^n|\ell;\,\, \ell>,
\end{equation}
and then
\begin{equation}
|\ell;\ell-2(n+1)>=\sqrt{\frac{(\ell-n-1)!}{(n-1)!\,\,\ell!}}L_{-}^{n+1}|\ell;\ell>.
\end{equation}
The action of the generator $L_{-}$ on the state $|\ell,\tilde{\ell};\ell-2n>$ is
\begin{eqnarray}
L_{-}|\ell; \ell-2n>&=&\sqrt{\frac{(\ell-n)!}{n!\,\,\ell!}} \,\,L_{-}^{n+1}|\ell;\ell>\nn \\
&=&\sqrt{\frac{(\ell-n-1)!}{(n+1)!\,\,\ell!}}\sqrt{(\ell-1)(n+1)} \,\,L_{-}^{n+1}|\ell;\ell>\nn \\
&=&\sqrt{(\ell-n)(n+1)} \,\,|\ell ;\ell-2(n+1)>.
\end{eqnarray}
The action of $L_{+}$ on $|\ell ;\ell -2n>$ yields
\begin{eqnarray}
L_{+}|\ell; \ell-2n>&=&\sqrt{\frac{(\ell-n)!}{n!\,\,\ell!}} \,\,L_{+}L_{-}^{n}|\ell; \ell>\nn \\
&=&\sqrt{\frac{(\ell-n)!}{n!\,\,\ell!}}L_{-}^nL_{+}+L_{-}^{n-1}[-n(R-n+1)]|\ell; \ell>\nn \\
&=&-\sqrt{\frac{(\ell-n+1)!}{(n-1)!\,\,\ell!}}\sqrt{(\ell-n+)(n)} \,\,L_{-}^{n+1}|\ell ;\ell> \nn\\
&=&-\sqrt{(\ell-n+1)(n)} \,\,|\ell;\ell-2(n-1)>.
\end{eqnarray}
From the above action of the generators $L_{-}$ and $L_{+}$, we have
\begin{eqnarray}
L_{-}|\ell;\ell-2n>&=&\sqrt{(\ell-n)(n+1)} \,\,|\ell;\ell-2(n+1)>, \\
L_{+}|\ell;\ell-2n>&=&-\sqrt{(\ell-n+1)(n)} \,\,|\ell;\ell-2(n-1)>.
\end{eqnarray}
Identifying $\ell-2n=m$ with consider $m$ as an even integer, we find $n=\frac{\ell-m}{2}$, $\ell-n=\frac{\ell+m}{2}$ and using this in the above relations we find
\begin{eqnarray}
L_{-}|\ell; m>&=&\hf \sqrt{(\ell+m)(\ell-m+2)} \,\,|\ell;\,\, m-2>, \\
L_{+}|\ell;\,\, m>&=&-\hf \sqrt{(\ell-m)(\ell+m+2)} \,\,|\ell;\,\, m+2>.
\end{eqnarray}
And if $\ell-2n=m$ with consider $m$ as an odd integer, corresponding to vectors in space $(0,1)$ or $(1,0)$, we can write $n=\frac{(\ell-1)-m}{2}$, $(\ell-1)-n=\frac{(\ell-1)+m}{2}$, yield
\begin{eqnarray}
L_{-}|\ell ; m>&=&\hf \sqrt{((\ell-1)+m)((\ell-1)-m+2)} \,\,|\ell ; m-2>,\\
L_{+}|\ell ;m>&=&-\hf \sqrt{((\ell-1)-m)((\ell-1)+m+2)} \,\,|\ell ; m+1>.
\end{eqnarray}
\noindent
 In order to find the action of $\tilde{L}_{-}$, we have to be careful, it is better to index the states 
 with the degree of the representation space, with the action :
 \begin{eqnarray}
 \tilde{L}_{-} |\ell ;\ell \rangle_{(0,0)}&=&|\ell ;\ell-2\rangle_{(1,1)}, \\
 \tilde{L}_{-} |\ell ;\ell \rangle_{(0,1)}&=&|\ell ;\ell-2\rangle_{(1,0)}, \\
  \tilde{L}_{-} |\ell  ;\ell \rangle_{(1,0)}&=&|\ell ;\ell-2\rangle_{(0,1)} ,\\
   \tilde{L}_{-} |\ell ;\ell \rangle_{(1,1)}&=&|\ell ;\ell-2\rangle_{(0,0)}. 
\end{eqnarray}
If the highest weight vector is in the sector $(0,0)$, we have the following results:
\begin{eqnarray}
|\ell; \ell-2n{\rangle}_{(0,0)} &=& \sqrt{\frac{(\ell-n)!}{n!\ell !}}{L_{-}}^n \,| \ell;\ell{\rangle}_{(0,0)},\\
L_{-} |\ell; \ell-2n{\rangle}_{(0,0)} &=&\sqrt{(\ell-n)(n+1)}\,\, |\ell\,\,;\,\,\ell-2n-2{\rangle}_{(0,0)}, \\
L_{+} |\ell; \ell-2n{\rangle}_{(0,0)} &=&-\sqrt{(\ell-n+1)n} \,\,|\ell \,\,;\,\,\ell-2n+2{\rangle}_{(0,0)}, \\
a_{-} |\ell; \ell-2n{\rangle}_{(0,0)} &=&\sqrt{2(\ell-n)}\,\, |\ell-1\,\,;\,\,\ell-2n-1{\rangle}_{(0,1)}, \\
a_{+} |\ell; \ell-2n{\rangle}_{(0,0)} &=&-\sqrt{2n}\,\, |\ell-1 \,\,; \,\, \ell-2n+1{\rangle}_{(0,1)}, \\
\tilde{a}_{-} |\ell; \ell-2n{\rangle}_{(0,0)} &=&\sqrt{2(\ell-n)}\,\, |\ell-1\,\,;\,\,\ell-2n-1{\rangle}_{(1,0)}, \\
\tilde{a}_{+} |\ell; \ell-2n{\rangle}_{(0,0)} &=&\sqrt{2n}\,\, |\ell-1 \,\,; \,\, \ell-2n+1{\rangle}_{(1,0)}, \\ 
L_{-}| \ell-1; \ell-2n-1 {\rangle}_{(0,1)}&=&\sqrt{(n+1)(\ell-1-n)}|\ell-1;\ell-2n-3{\rangle}_{(0,1)}, \\
L_{+} |\ell-1; \ell-2n-1 {\rangle}_{(0,1)}&=&-\sqrt{n(\ell-1-n+1)}|\ell-1;\ell-2n+1{\rangle}_{(0,1)}, \\
a_{-}|\ell-1;\ell-2n-1{\rangle}_{(0,1)}&=&\sqrt{2(n+1)}|\ell;\ell-2n-2{\rangle}_{(0,0)},\\
a_{+}|\ell-1;\ell-2n-1{\rangle}_{(0,1)}&=&\sqrt{2(\ell-n)}|\ell;\ell-2n {\rangle}_{(0,0)}, \\
\tilde{a}_{-}|\ell-1;\ell-1-2n{\rangle}_{(0,1)}&=&-\sqrt{2(n+1)}|\ell;\ell-2n-2{\rangle}_{(1,1)}, \\
\tilde{a}_{+}|\ell-1;\ell-2n-1{\rangle}_{(0,1)}&=&\sqrt{2(\ell-n)}|\ell;\ell-2n{\rangle}_{(1,1)}, \\
L_{-}| \ell-1; \ell-2n-1 {\rangle}_{(1,0)}&=&\sqrt{(n+1)(\ell-1-n)}|\ell-1;\ell-2n-3{\rangle}_{(1,0)}, \\
L_{+} |\ell-1; \ell-2n-1 {\rangle}_{(1,0)}&=&-\sqrt{n(\ell-1-n+1)}|\ell-1;\ell-2n+1{\rangle}_{(1,0)}, \\
a_{-}|\ell-1;\ell-2n-1{\rangle}_{(1,0)}&=&\sqrt{2(n+1)}|\ell;\ell-2n-2{\rangle}_{(1,1)}, \\
a_{+}|\ell-1;\ell-2n-1{\rangle}_{(1,0)}&=&\sqrt{2(\ell-n)}|\ell;\ell-2n{\rangle}_{(1,1)}, \\
\tilde{a}_{-} |\ell-1;\ell-2n-1{\rangle}_{(1,0)}&=&-\sqrt{2(n+1)}|\ell;\ell-2n-2{\rangle}_{(0,0)}, \\
\tilde{a}_{+}|\ell-1;\ell-2n-1{\rangle}_{(1,0)}&=&\sqrt{2(\ell-n)}|\ell;\ell-2n{\rangle}_{(0,0)}, \\
L_{-} |\ell; \ell-2n{\rangle}_{(1,1)} &=&\sqrt{(\ell-n)(n+1)}\,\, |\ell\,\,;\,\,\ell-2n-2{\rangle}_{(1,1)}, \\
L_{+} |\ell; \ell-2n{\rangle}_{(1,1)} &=&-\sqrt{(\ell-n+1)n} \,\,|\ell \,\,;\,\,\ell-2n+2{\rangle}_{(1,1)}, \\
a_{-} |\ell; \ell-2n{\rangle}_{(1,1)} &=&\sqrt{2(\ell-n)}\,\, |\ell-1\,\,;\,\,\ell-2n-1{\rangle}_{(1,0)}, \\
a_{+} |\ell; \ell-2n{\rangle}_{(1,1)} &=&-\sqrt{2n}\,\, |\ell-1 \,\,; \,\, \ell-2n+1{\rangle}_{(1,0)}, \\
\tilde{a}_{-} |\ell; \ell-2n{\rangle}_{(1,1)} &=&\sqrt{2(\ell-n)}\,\, |\ell-1\,\,;\,\,\ell-2n-1{\rangle}_{(0,1)}, \\
\tilde{a}_{+} |\ell; \ell-2n{\rangle}_{(1,1)} &=&\sqrt{2n}\,\, |\ell-1 \,\,; \,\, \ell-2n+1{\rangle}_{(0,1)}. 
\end{eqnarray}
 Using the above, taking the highest weight vector in $(0,0)$ sector and $\ell=2$ for the highest weight vector
 we have the following basis
\begin{equation}
\begin{array}{cccc}
(0,0) : &|2,2{\rangle}_{(0,0)},&|2,0{\rangle}_{(0,0)},&|2,-2{\rangle}_{(0,0)}; \\
(0,1) : &|1,1{\rangle}_{(0,1)},&|1,-1{\rangle}_{(0,1)}; & \\
(1,0) : &|1,1{\rangle}_{(1,0)},&|1,-1{\rangle}_{(1,0)}; & \\
(1,1) : &|2,2{\rangle}_{(1,1)},&|2,0{\rangle}_{(1,1)},&|2,-2{\rangle}_{(1,1)}. 
\end{array}
\end{equation} 
\section{Matrix form}
The above basis provide a nontrivial fundamental irreducible  basis of states for the ten generator algebra. Using the general results of the action of the generators on the states, one  can represent the generators $L_{+},L_{-},...$ as $ 10 \, \times \, 10$ matrix form in the following:
\begin{equation}
L_{+}=
\begin{bmatrix}
0 & -\sqrt{2} & 0 & 0 & 0 & 0 & 0 & 0 & 0 & 0 \\
0 & 0 & -\sqrt{2} & 0 & 0 & 0 & 0 & 0 & 0 & 0 \\
0 & 0 & 0 & 0 & 0 & 0 & 0 & 0 & 0 & 0 \\
0 & 0 & 0 & 0 & -1 & 0 & 0 & 0 & 0 & 0 \\
0 & 0 & 0 & 0 & 0 & 0 & 0 & 0 & 0 & 0 \\
0 & 0 & 0 & 0 & 0  & 0 & -1 & 0 & 0 & 0 \\
0 & 0 & 0 & 0 & 0 & 0 & 0 & 0 & 0 & 0 \\
0 & 0 & 0 & 0 & 0  & 0 & 0 & 0 & -\sqrt{2} & 0 \\
0 & 0 & 0 & 0 & 0  & 0 & 0 & 0 & 0 & -\sqrt{2} \\
0 & 0 & 0 & 0 & 0 & 0 & 0 & 0 & 0 & 0 
\end{bmatrix}
\end{equation}

\begin{equation}
L_{-}=
\begin{bmatrix}
0 & 0 & 0 & 0 & 0 & 0 & 0 & 0 & 0 & 0 \\
\sqrt{2} & 0 & 0 & 0 & 0 & 0 & 0 & 0 & 0 & 0 \\
0 &\sqrt{2} & 0 & 0 & 0 & 0 & 0 & 0 & 0 & 0 \\
0 & 0 & 0 & 0 & 0 & 0 & 0 & 0 & 0 & 0 \\
0 & 0 & 0 & 1 & 0 & 0 & 0 & 0 & 0 & 0 \\
0 & 0 & 0 & 0 & 0  & 0 & 0 & 0 & 0 & 0 \\
0 & 0 & 0 & 0 & 0 & 1 & 0 & 0 & 0 & 0 \\
0 & 0 & 0 & 0 & 0  & 0 & 0 & 0 & 0 & 0 \\
0 & 0 & 0 & 0 & 0  & 0 & 0 & \sqrt{2} & 0 & 0 \\
0 & 0 & 0 & 0 & 0 & 0 & 0 & 0 & \sqrt{2} & 0 
\end{bmatrix}
\end{equation}

\begin{equation}
a_{+}=
\begin{bmatrix}
0 & 0 & 0 & 2 & 0 & 0 & 0 & 0 & 0 & 0 \\
0 & 0 & 0 & 0 & \sqrt{2} & 0 & 0 & 0 & 0 & 0 \\
0 & 0 & 0 & 0 & 0 & 0 & 0 & 0 & 0 & 0 \\
0 & -\sqrt{2} & 0 & 0 & 0 & 0 & 0 & 0 & 0 & 0 \\
0 & 0 & -2 & 0 & 0 & 0 & 0 & 0 & 0 & 0 \\
0 & 0 & 0 & 0 & 0  & 0 & 0 & 0 & -\sqrt{2} & 0 \\
0 & 0 & 0 & 0 & 0 & 0 & 0 & 0 & 0 & -2 \\
0 & 0 & 0 & 0 & 0  & 2 & 0 & 0 & 0 & 0 \\
0 & 0 & 0 & 0 & 0  & 0 & \sqrt{2} & 0 & 0 & 0\\
0 & 0 & 0 & 0 & 0 & 0 & 0 & 0 & 0 & 0 
\end{bmatrix}
\end{equation}

\begin{equation}
a_{-}=
\begin{bmatrix}
0 & 0 & 0 & 0 & 0 & 0 & 0 & 0 & 0 & 0 \\
0 & 0 & 0 & \sqrt{2} & 0 & 0& 0 & 0 & 0 & 0 \\
0 & 0 & 0 & 0 & 2 & 0 & 0 & 0 & 0 & 0 \\
2 & 0 & 0 & 0 & 0 & 0 & 0 & 0 & 0 & 0 \\
0 & \sqrt{2} & 0 & 0 & 0 & 0 & 0 & 0 & 0 & 0 \\
0 & 0 & 0 & 0 & 0  & 0 & 0 & 2 & 0 & 0 \\
0 & 0& 0 & 0 & 0 & 0 & 0 & 0 & \sqrt{2} &0 \\
0 & 0 & 0 & 0 & 0  & 0 & 0 & 0 & 0 & 0 \\
0 & 0 & 0 & 0 & 0  & \sqrt{2} & 0& 0 & 0 & 0\\
0 & 0 & 0 & 0 & 0 & 0 & 2 & 0 & 0 & 0 
\end{bmatrix}
\end{equation}

\begin{equation}
\tilde{a}_{+}=
\begin{bmatrix}
0 & 0 & 0 & 0 & 0 & 2 & 0 & 0 & 0 & 0 \\
0 & 0 & 0 & 0 & 0 & 0 & \sqrt{2} & 0 & 0 & 0 \\
0 & 0 & 0 & 0 & 0 & 0 & 0 & 0 & 0 & 0 \\
0 & 0 & 0 & 0 & 0 & 0 & 0 & 0 & \sqrt{2} & 0 \\
0 & 0 & 0 & 0 & 0 & 0 & 0 & 0 & 0 & 2\\
0 & \sqrt{2} & 0 & 0 & 0  & 0 & 0 & 0 & 0 & 0 \\
0 & 0 & 2 & 0 & 0 & 0 & 0 & 0 & 0 & 0\\
0 & 0 & 0 & 2 & 0  & 0 & 0 & 0 & 0 & 0 \\
0 & 0 & 0 & 0& \sqrt{2} & 0 & 0 & 0 & 0 & 0\\
0 & 0 & 0 & 0 & 0 & 0 & 0 & 0 & 0 & 0 
\end{bmatrix}
\end{equation}

\begin{equation}
\tilde{a}_{-}=
\begin{bmatrix}
0 & 0 & 0 & 0 & 0 & 0 & 0 & 0 & 0 & 0 \\
0 & 0 & 0 & 0 & 0 & -\sqrt{2} & 0 & 0 & 0 & 0 \\
0 & 0 & 0 & 0 & 0 & 0 & -2 & 0 & 0 & 0 \\
0 & 0 & 0 & 0 & 0 & 0 & 0 & 2 & 0 & 0 \\
0 & 0 & 0 & 0 & 0 & 0 & 0 & 0 & \sqrt{2} & 0 \\
2 & 0 & 0 & 0 & 0  & 0 & 0 & 0 & 0 & 0 \\
0 & \sqrt{2}& 0 & 0 & 0 & 0 & 0 & 0 & 0 &0 \\
0 & 0 & 0 & 0 & 0  & 0 & 0 & 0 & 0 & 0 \\
0 & 0 & 0 & -\sqrt{2} & 0  & 0 & 0& 0 & 0 & 0\\
0 & 0 & 0 & 0 & -2 & 0 & 0 & 0 & 0 & 0 
\end{bmatrix}
\end{equation}

\begin{equation}
{R}=
\begin{bmatrix}
2 & 0 & 0 & 0 & 0 & 0 & 0 & 0 & 0 & 0 \\
0 & 0 & 0 & 0 & 0 & 0 & 0 & 0 & 0 & 0 \\
0 & 0 & -2 & 0 & 0 & 0 & 0 & 0 & 0 & 0 \\
0 & 0 & 0 & 1 & 0 & 0 & 0 & 0 & 0 & 0 \\
0 & 0 & 0 & 0 & -1 & 0 & 0 & 0 & 0 & 0 \\
0& 0 & 0 & 0 & 0  & 1 & 0 & 0 & 0 & 0 \\
0 & 0 & 0 & 0 & 0 & 0 & -1 & 0 & 0 &0 \\
0 & 0 & 0 & 0 & 0  & 0 & 0 & 2 & 0 & 0 \\
0 & 0 & 0 & 0  & 0  & 0 & 0& 0 & 0 & 0\\
0 & 0 & 0 & 0 & 0 & 0 & 0 & 0 & 0 & -2 
\end{bmatrix}
\end{equation}

\begin{equation}
\tilde{R}=
\begin{bmatrix}
0 & 0 & 0 & 0 & 0 & 0 & 0 & 2 & 0 & 0 \\
0 & 0 & 0 & 0 & 0 & 0 & 0 & 0 & 0 & 0 \\
0 & 0 & 0 & 0 & 0 & 0 & 0 & 0 & 0 & -2 \\
0 & 0 & 0 & 0 & 0 & -1 & 0 & 0 & 0 & 0 \\
0 & 0 & 0 & 0 & 0 & 0 & 1 & 0 & 0 & 0 \\
0& 0 & 0 & -1 & 0  & 0 & 0 & 0 & 0 & 0 \\
0 & 0 & 0 & 0 & 1 & 0 & 0 & 0 & 0 &0 \\
2 & 0 & 0 & 0 & 0  & 0 & 0 & 0 & 0 & 0 \\
0 & 0 & 0 & 0  & 0  & 0 & 0& 0 & 0 & 0\\
0 & 0 & -2 & 0 & 0 & 0 & 0 & 0 & 0 & 0 
\end{bmatrix}
\end{equation}

\begin{equation}
\tilde{L}_{+}=
\begin{bmatrix}
0 & 0 & 0 & 0 & 0 & 0 & 0 & 0 & -\sqrt{2} & 0 \\
0 & 0 & 0 & 0 & 0 & 0 & 0 & 0 & 0 &-\sqrt{2} \\
0 & 0 & 0 & 0 & 0 & 0 & 0 & 0 & 0 & 0 \\
0 & 0 & 0 & 0 & 0 & 0 & 1 & 0 & 0 & 0 \\
0 & 0 & 0 & 0 & 0 & 0 & 0 & 0 & 0 & 0 \\
0 & 0 & 0 & 0 & 1 & 0 & 0 & 0 & 0 & 0 \\
0 & 0 & 0 & 0 & 0 & 0 & 0 & 0 & 0 & 0 \\
0 & -\sqrt{2}& 0 & 0 & 0  & 0 & 0 & 0 & 0 & 0 \\
0 & 0 &-\sqrt{2}& 0 & 0  & 0 & 0 & 0 & 0 & 0\\
0 & 0 & 0 & 0 & 0 & 0 & 0 & 0 & 0 & 0 
\end{bmatrix}
\end{equation}
\begin{equation}
\tilde{L}_{-}=
\begin{bmatrix}
0 & 0 & 0 & 0 & 0 & 0 & 0 & 0 & 0 & 0 \\
0 & 0 & 0 & 0 & 0 & 0 & 0 & \sqrt{2} & 0 & 0 \\
0 &0 & 0 & 0 & 0 & 0 & 0 & 0 &\sqrt{2} & 0 \\
0 & 0 & 0 & 0 & 0 & 0 & 0 & 0 & 0 & 0 \\
0 & 0 & 0 & 0 & 0 & -1 & 0 & 0 & 0 & 0 \\
0 & 0 & 0 & 0 & 0  & 0 & 0 & 0 & 0 & 0 \\
0 & 0 & 0 & -1 & 0 & 0 & 0 & 0 & 0 & 0 \\
0 & 0 & 0 & 0 & 0  & 0 & 0 & 0 & 0 & 0 \\
\sqrt{2} & 0 & 0 & 0 & 0  & 0 & 0 & 0 & 0 & 0 \\
0 & \sqrt{2} & 0 & 0 & 0 & 0 & 0 & 0 & 0 & 0 
\end{bmatrix}
\end{equation}
It is a an easy exercise to check that the above matrices satisfy the defining algebra, and substituting it in the Casimir gives a multiple of identity operator, proving its irreducibility.
\section{Alternative algebra}
It is found in \cite{RY} that there are two independent versions for the \Z2 extensions of the $osp(1|2)$, the one that we studied in the previous sections and another one with eight generators that we study in below with the algebra
\begin{equation}
\begin{array}{llll}
 [R,L_{\pm}] = \pm 2 L_{\pm}, &[L_{+},L_{-}] = -R ,& [R, a_{\pm}] = \pm a_{\pm}, & 
 [ R, {\tilde{a}}_{\pm}] = \pm {\tilde{a}}_{\pm}, \\[3pt]
[\tilde{R},L_{\pm}] = 0, &[L_{\pm}, a_{\mp}] = \mp a_{\pm} ,& 
\{\tilde{R}, a_{\pm}\} = \pm \tilde{a}_{\pm}, & \{\tilde{R}, \tilde{a}_{\pm}\} = \mp a_{\pm}, \\[3pt]
\{a_{+}, a_{-}\} = 2R, & [a_{\pm}, \tilde{a}_{\mp}] = 2 \tilde{R},& \{ \tilde{a}_{-}, \tilde{a}_{+} \} = 2R, &
[L_{\pm}, \tilde{a}_{\mp}] = \pm \tilde{a}_{\pm}, \\[3pt]
[a_{\pm}, \tilde{a}_{\pm}] = 0, & \{a_{\pm}, a_{\pm}\} = 4L_{\pm},& 
\{\tilde{a}_{\pm}, \tilde{a}_{\pm}\} = -4L_{\pm}. &
\end{array}
\label{Z22}
\end{equation}
This is essentially the same algebra in \eqref{Z21}, thus the Casimir operator for this algebra can be easily 
found, that is,
\begin{equation}
C_2=\alpha R + \beta R^2 + \gamma \tilde{R}^2 +\delta L_{+}L_{-} +\rho a_{+}a_{-} + \sigma \tilde{a}_{+}\tilde{a}_{-} .
\end{equation}
Fixing the constants, we may rewrite the Casimir invariant
\begin{equation}
C_2=\alpha (-\frac{1}{4} R^2 -\frac{1}{4}  \tilde{R}^2 + L_{+}L_{-} +\frac{1}{4} a_{+}a_{-} + \frac{1}{4} \tilde{a}_{+}\tilde{a}_{-}) + \beta,
\end{equation}
where $\alpha$ and $\beta$ are any arbitrary constants.
This algebra does not have a quadratic Casimir in $(1,1)$ sector.
However, we may find a vector field realization for the algebra in terms of parameters with the grading
\begin{eqnarray}
 \theta &:&  \in (0,1), \\ \psi &:&  \in (1,0), \\ z  &:&  \in (0,0).
\end{eqnarray}
In this space, the realization of the algebra provides us the following first-order differential generators
\begin{eqnarray}
a_{-}&=&\partial_{\theta}+2 \theta \partial_z, \\[3pt]
a_{+}&=&z\partial_{\theta}-2\psi \tilde{r} +2\theta r + 2\theta z \partial_z + 2\theta\psi \partial_{\psi}, \\[3pt]
R&=&r+\theta \partial_{\theta}+\psi \partial_{\psi} +2z\partial_z,\\[3pt]
\tilde{a}_{-}&=&\partial_{\psi}-2\psi \partial_{z}, \\[3pt]
\tilde{a}_{+}&=&-z\partial_{\psi}+2\theta \tilde{r} +2\psi r + 2\psi z \partial_z + 2\theta\psi \partial_{\theta}, \\[3pt]
\tilde{R}&=&\tilde{r}+\psi \partial_{\theta}-\theta \partial_{\psi}, \\[3pt]
L_{-}&=&\partial_z, \\[3pt]
L_{+}&=&z r+z^2\partial_z+z\psi \partial_{\psi}+z\theta \partial_{\theta}-2\theta \psi \tilde{r}.
\end{eqnarray}
We now work out the irreducible representations for the eight generators case with the basis vectors are:
\begin{equation}
\begin{array}{cccc}
(0,0): &|2,2{\rangle}_{(0,0)},&|2,0{\rangle}_{(0,0)},&|2,-2{\rangle}_{(0,0)}; \\
(0,1): &|1,1{\rangle}_{(0,1)},&|1,-1{\rangle}_{(0,1)};& \\
(1,0): &|1,1{\rangle}_{(1,0)},&|1,-1{\rangle}_{(1,0)};& \\
(1,1): &|0,0{\rangle}_{(1,1)}& 
\end{array}
\end{equation} 
and their corresponding matrix representations
\begin{equation}
L_{+}=
\begin{bmatrix}
0 & -\sqrt{2} & 0 & 0 & 0 & 0 & 0 & 0 \\
0 & 0 & -\sqrt{2} & 0 & 0 & 0 & 0 & 0  \\
0 & 0 & 0 & 0 & 0 & 0 & 0 & 0  \\
0 & 0 & 0 & 0 & -1 & 0 & 0 & 0  \\
0 & 0 & 0 & 0 & 0 & 0 & 0 & 0  \\
0 & 0 & 0 & 0 & 0  & 0 & -1 & 0 \\
0 & 0 & 0 & 0 & 0 & 0 & 0 & 0  \\
0 & 0 & 0 & 0 & 0  & 0 & 0 & 0  \\
\end{bmatrix}
\end{equation}

\begin{equation}
L_{-}=
\begin{bmatrix}
0 & 0 & 0 & 0 & 0 & 0 & 0 & 0  \\
\sqrt{2} & 0 & 0 & 0 & 0 & 0 & 0 & 0  \\
0 &\sqrt{2} & 0 & 0 & 0 & 0 & 0 & 0  \\
0 & 0 & 0 & 0 & 0 & 0 & 0 & 0  \\
0 & 0 & 0 & 1 & 0 & 0 & 0 & 0 \\
0 & 0 & 0 & 0 & 0  & 0 & 0 & 0  \\
0 & 0 & 0 & 0 & 0 & 1 & 0 & 0\\
0 & 0 & 0 & 0 & 0  & 0 & 0 & 0  \\
\end{bmatrix}
\end{equation}

\begin{equation}
a_{+}=
\begin{bmatrix}
0 & 0 & 0 & 2 & 0 & 0 & 0 & 0 \\
0 & 0 & 0 & 0 & \sqrt{2} & 0 & 0 & 0  \\
0 & 0 & 0 & 0 & 0 & 0 & 0 & 0  \\
0 & -\sqrt{2} & 0 & 0 & 0 & 0 & 0 & 0 \\
0 & 0 & -2 & 0 & 0 & 0 & 0 & 0  \\
0 & 0 & 0 & 0 & 0  & 0 & 0 & \sqrt{2} \\
0 & 0 & 0 & 0 & 0 & 0 & 0 & 0 \\
0 & 0 & 0 & 0 & 0  & 0 & -\sqrt{2} & 0  
\end{bmatrix}
\end{equation}

\begin{equation}
a_{-}=
\begin{bmatrix}
0 & 0 & 0 & 0 & 0 & 0 & 0 & 0 \\
0 & 0 & 0 & \sqrt{2} & 0 & 0& 0 & 0 \\
0 & 0 & 0 & 0 & 2 & 0 & 0 & 0 \\
2 & 0 & 0 & 0 & 0 & 0 & 0 & 0 \\
0 & \sqrt{2} & 0 & 0 & 0 & 0 & 0 & 0 \\
0 & 0 & 0 & 0 & 0  & 0 & 0 & 0 \\
0 & 0& 0 & 0 & 0 & 0 & 0 & \sqrt{2} \\
0 & 0 & 0 & 0 & 0  & \sqrt{2} & 0 & 0 
\end{bmatrix}
\end{equation}

\begin{equation}
\tilde{a}_{+}=
\begin{bmatrix}
0 & 0 & 0 & 0 & 0 & 2 & 0 & 0 \\
0 & 0 & 0 & 0 & 0 & 0 & \sqrt{2} & 0  \\
0 & 0 & 0 & 0 & 0 & 0 & 0 & 0  \\
0 & 0 & 0 & 0 & 0 & 0 & 0 & \sqrt{2} \\
0 & 0 & 0 & 0 & 0 & 0 & 0 & 0 \\
0 & \sqrt{2} & 0 & 0 & 0  & 0 & 0 & 0  \\
0 & 0 & 2 & 0 & 0 & 0 & 0 & 0\\
0 & 0 & 0 & 0 & \sqrt{2} & 0 & 0 & 0  
\end{bmatrix}
\end{equation}

\begin{equation}
\tilde{a}_{-}=
\begin{bmatrix}
0 & 0 & 0 & 0 & 0 & 0 & 0 & 0 \\
0 & 0 & 0 & 0 & 0 & -\sqrt{2} & 0 & 0 \\
0 & 0 & 0 & 0 & 0 & 0 & -2 & 0 \\
0 & 0 & 0 & 0 & 0 & 0 & 0 & 0 \\
0 & 0 & 0 & 0 & 0 & 0 & 0 & -\sqrt{2} \\
2 & 0 & 0 & 0 & 0  & 0 & 0 & 0  \\
0 & \sqrt{2}& 0 & 0 & 0 & 0 & 0 & 0 \\
0 & 0 & 0 & \sqrt{2} & 0  & 0 & 0 & 0 \\
\end{bmatrix}
\end{equation}

\begin{equation}
{R}=
\begin{bmatrix}
2 & 0 & 0 & 0 & 0 & 0 & 0 & 0  \\
0 & 0 & 0 & 0 & 0 & 0 & 0 & 0  \\
0 & 0 & -2 & 0 & 0 & 0 & 0 & 0  \\
0 & 0 & 0 & 1 & 0 & 0 & 0 & 0 \\
0 & 0 & 0 & 0 & -1 & 0 & 0 & 0 \\
0& 0 & 0 & 0 & 0  & 1 & 0 & 0  \\
0 & 0 & 0 & 0 & 0 & 0 & -1 & 0\\
0 & 0 & 0 & 0 & 0  & 0 & 0 & 0\\
\end{bmatrix}
\end{equation}

\begin{equation}
\tilde{R}=
\begin{bmatrix}
0 & 0 & 0 & 0 & 0 & 0 & 0 & 0 \\
0 & 0 & 0 & 0 & 0 & 0 & 0 & 0  \\
0 & 0 & 0 & 0 & 0 & 0 & 0 & 0 \\
0 & 0 & 0 & 0 & 0 & 1 & 0 & 0  \\
0 & 0 & 0 & 0 & 0 & 0 & 1 & 0 \\
0& 0 & 0 & -1 & 0  & 0 & 0 & 0  \\
0 & 0 & 0 & 0 & -1 & 0 & 0 & 0\\
0 & 0 & 0 & 0 & 0  & 0 & 0 & 0  
\end{bmatrix}
\end{equation}

For these representations, the Casimir becomes a multiple of the identity.

\section{Conclusion}\label{sec6}
In this work, we constructed the finite-dimensional irreducible representations of the \Z2 extended  $osp(1|2)$ algebra. We note that two different versions of $osp(1|2)$ \Z2 extended algebras exist, one with ten generators and another with eight generators.  
We construct the quadratic Casimir for the \Z2 extended $osp(1|2)$ algebra, and it is found that the ten-generator version of the algebra has two quadratic Casimirs one in the $(0,0)$ sector and one in the $(1,1)$ sector. We deduced first-order differential operator realization of both versions of the algebras. We note that the differential operator realization substituted in the Casimir provides an identity operator, which proves its irreducibility. We hope this differential operator realization will find use in the applications of these algebras. We explicitly work out the finite-dimensional irreducible representation of these algebras starting from the highest weight vectors. We use these vectors to calculate the matrix representations for the generators of the algebras. We are working on the applications of these representations we have found here.

\section*{Acknowledgement}
FH was partially supported by the project grant CZ.02.2.69/0.0/0.0/18\_053/0016980 Mobility CTU - STA, Ministry of Education, Youth and Sports of the Czech Republic, co-financed by the European Union. The authors thank N. Aizawa (Department of Physics, Graduate School of Science, Osaka Metropolitan University, Japan) for constructive discussions and helpful comments on the subject of this manuscript.


\begin{thebibliography}{}


\bibitem{rw1} V. Rittenberg and D. Wyler, Generalized Superalgebras, Nucl. Phys. {\bf B 139} 189 (1978)

\bibitem{rw2} V. Rittenberg and D. Wyler, Sequences of $Z_2\otimes Z_2$ graded Lie algebras and superalgebras, J. Math. Phys. {\bf 19}  2193 (1978)

\bibitem{Ree1960} R. Ree, Generalized Lie elements, Canad. J. Math. \textbf{12} 493 (1960)

\bibitem{Sch1979} M. Scheunert, Generalized Lie algebras, J. Math. Phys. {\bf 20} 712 (1979)

\bibitem{Zha2003}
K. Zhao, Simple Lie color algebras from graded associative algebras, Journal of Algebra, {\bf 269}, 439 (2003).
%


\bibitem{aktt1} N. Aizawa, Z. Kuznetsova, H. Tanaka and F. Toppan, \Z2-graded Lie symmetries of the L\'evy-Leblond equations, Prog. Theor. Exp. Phys. \textbf{2016} 123A01 (2016)

\bibitem{NaPsiJs} N. Aizawa, P. S. Isaac and J. Segar, $\mathbb{Z}_2 \times \mathbb{Z}_2$
generalizations of $ {\cal N}=1 $ superconformal Galilei algebras and their representations, J.
Math. Phys. \textbf{60}, 023507 (2019)

\bibitem{NA2} N. Aizawa, Generalization of superalgebras to color superalgebras and their
representations, Adv. Appl. Clifford Algebras \textbf{28}, 28 (2018)


\bibitem{AzKaDo} N. Aizawa, K. Amakawa  and S. Doi, $\mathcal{N}$-extension of double-graded supersymmetric and superconformal quantum mechanics, J. Phys. A: Math. Theor. { \bf 53}, 065205 (2020)
 
\bibitem{Dee85} J. Deenen and C. Quesne, Boson representations of the real symplectic group
and their applications to the nuclear collective model, J. Math. Phys. {\bf 26}, 2705 (1985)

\bibitem{Fil81}
G. F. Filippov, V.I. Ovcharenko and Y. F. Smirnov, Microscopic Theory of Collective Excitations of Atomic Nuclei, Naukova Dumka, Kiev (Russian) (1981)
%
 
\bibitem{Iac80}
F. Iachello, Dynamical supersymmetries in nuclei, Phys. Rev. Lett., {\bf 44}, 772 (1980)

\bibitem{PhJo} P. S. Issac, N. I. Stoilova and J. V. der Jeugt, The \Z2 - graded general linear Lie superalgebra, 
J. Math. Phys. {\bf 61}, 011702 (2020)

\bibitem{AzKa} K. Amakawa and N. Aizawa, A classification of lowest weight irreducible modules over
$\mathbb{Z}^2_2$-graded extension of $osp(1|2)$, J. Math. Phys. {\bf 62}, 043502, (2021) 

\bibitem{Fa} 
F. S. Alshammari, Phillip S. Isaac, Ian Marquettee, A differential operator realisation approach for constructing Casimir operators of non-semisimple Lie algebras, J. Phys.{\bf A51}, 065206 (2018)
%

\bibitem{Caosp12} N. Backhouse, Computing invariants of Lie Superalgebras, J. Phys. A: Math. Gen, {\bf 12}, 21 (1979) 


\bibitem{RY} R. Lu and Y. Tan, Construction of color Lie algebras from homomorphisms of modules of Lie algebras, J. Alg. {\bf 620} 1-49 (2023)

\end{thebibliography}
\end{document}